\documentclass[twocolumn,tighten]{aastex62}

\usepackage{amsmath}
\usepackage{verbatim}
\shorttitle{Dust distribution of the Milky Way}
\shortauthors{Guo et al.}

\begin{document}

\title{Three-Dimensional Distribution of the Interstellar Dust in the Milky Way}

\author{H.-L. Guo}
\affiliation{South-Western Institute for Astronomy Research, Yunnan University, Kunming 650500, People's Republic of China}

\author{B.-Q. Chen}
\thanks{Corresponding authors: B.-Q. Chen (bchen$@$ynu.edu.cn; BQC)\\ and X.-W. Liu (x.liu$@$ynu.edu.cn; XWL)}
\affil{South-Western Institute for Astronomy Research, Yunnan University, Kunming 650500, People's Republic of China}

\author{H.-B. Yuan}
\affil{Department of Astronomy, Beijing Normal University, Beijing 100875, People's Republic of China}

\author{ Y. Huang}
\affil{South-Western Institute for Astronomy Research, Yunnan University, Kunming 650500, People's Republic of China}

\author{D.-Z Liu}
\affil{South-Western Institute for Astronomy Research, Yunnan University, Kunming 650500, People's Republic of China}

\author{Y. Yang}
\affil{South-Western Institute for Astronomy Research, Yunnan University, Kunming 650500, People's Republic of China}

\author{X.-Y. Li}
\affil{South-Western Institute for Astronomy Research, Yunnan University, Kunming 650500, People's Republic of China}

\author{W.-X. Sun}
\affil{South-Western Institute for Astronomy Research, Yunnan University, Kunming 650500, People's Republic of China}

\author{X.-W. Liu}
\thanks{Corresponding authors: B.-Q. Chen (bchen$@$ynu.edu.cn; BQC)\\ and X.-W. Liu (x.liu$@$ynu.edu.cn; XWL)}
\affil{South-Western Institute for Astronomy Research, Yunnan University, Kunming 650500, People's Republic of China}



\begin{abstract}

We present a three-dimensional (3D) extinction map of the southern sky. The map covers the SkyMapper Southern Survey (SMSS) area of $\sim$ 14,000\ ${\rm deg^{2}}$ and has spatial resolutions between 6.9 and 27\,arcmin. Based on the multi-band photometry of SMSS, the Two Micron All Sky Survey, the Wide-Field Infrared Survey Explorer Survey and the Gaia mission, we have estimated values of the $r$-band extinction for $\sim$ 19 million stars with the spectral energy distribution (SED) analysis. Together with the distances calculated from the Gaia data release 2 (DR2) parallaxes, we have constructed a three-dimensional extinction map of the southern sky. By combining our 3D extinction map with those from the literature, we present an all-sky 3D extinction map, and use it to explore the 3D distribution of the Galactic dust grains. We use two different models, one consisting a single disk and an other of two disks, to fit the 3D distribution of the Galactic dust grains. The data is better fitted by a two-disk model, yielding smaller values of the Bayesian Information Criterion (BIC). The best fit model has scale heights of 73 and 225\,pc for the ``thin" and ``thick" dust disks, respectively.


\end{abstract}




\section{Introduction} \label{sec:intro}

Extinction and reddening by the interstellar dust grains pose a serious obstacle for the study of the structure and stellar populations of the Milky Way galaxy. In order to obtain the intrinsic luminosities or colors of the observed objects, one needs to correct for the dust extinction and reddening. Extinction maps are useful tools for this purpose. The traditional two-dimensional (2D) extinction maps, including those from dust emission in far-infrared (IR; \citealt{Schlegel1998}; hereafter SFD), far-infrared combined with microwave \citep{Planck2014}, and those derived from optical and near-IR stellar photometry \citep [e.g.][]{Schlafly2010, Majewski2011, Nidever2012, Gonzalez2012, Gonzalez2018}, give only the total or an average extinction for a given line of sight and therefore do not deliver the information of dust distribution as a function of distance. This can be particularly problematic for Galactic objects at a finite distance, especially those in the disk.

To overcome the limitation of the 2D extinction maps, several attempts have been carried out to build three-dimensional (3D) extinction maps based on data from recent large-scale photometric, spectroscopic and astrometric surveys. Depending
on the adopted data sets, those 3D extinction maps have different sky coverages. For example, \citet{Marshall2006} presented a 3D extinction map of the inner Galaxy ($|l|$ $<$ 100\degr\ and $|b|$ $<$ 10\degr) based on data from the Two Micron All Sky Survey \citep[2MASS;][] {Skrutskie2006}. \citet{Chen2013} and \citet{Schultheis2014} presented maps toward the Galactic Bulge ($|l|$ $<$ 10\degr\ and $-$10\degr $<$ $b$ $<$ 5\degr) based on data of the Galactic Legacy Infrared Mid-Plane Survey Experiment \citep[GLIMPSE;][]{Churchwell2009}, the VISTA Variables in the Via Lactea Survey \citep[VVV;][] {Minniti2010} and 2MASS. \citet{Chen2014} presented maps toward the Galactic anti-center (140\degr $<$ $l$ $<$ 220\degr\ and $-$60\degr $<$ $b$ $<$ 40\degr) based on data from the Xuyi Schmidt Telescope Photometric Survey of the Galactic Anticentre \citep[XSTPS-GAC;][]{2014IAUS}, 2MASS and Wide-Field Infrared Survey Explorer \citep[WISE;][]{Wright2010}. \citet{Sale2014} presented maps of the Northern Galactic Plane (30\degr $<$ $l$ $<$ 215\degr\ and $| b |$ $<$ 5\degr) based on data from the INT/WFC Photometric H$\alpha$ Survey \citep[IPHAS;][]{Drew2005}. \citet{Green2015}, \citet{Green2018} and \citet{Green2019} presented maps of the northern sky ($\delta$ $>$ $-$30\degr) based on data from the Pan-STARRS 1 Survey \citep[PS1;][] {Chambers2016}, 2MASS and Gaia DR2 \citep{Gaia2016, Gaia2018}. \citet{Chen2019} and \citet{Lallement2019} presented maps of the Galactic mid-plane based on data from Gaia DR2, 2MASS and WISE.

As most of the recently completed large-scale surveys  target the northern sky, most of the above 3D extinction maps have coverages only in the northern parts of the sky. A 3D extinction map for the southern sky is still missing and should be valuable to have in order to explore the global dust distribution of the Milky Way. The first data release of the SkyMapper Southern Survey \citep[SMSS DR1;][]{Wolf2018} provide us with the opportunity to fill this gap. In this paper, we combine the optical photometry of SMSS DR1 with the infrared (IR) photometry of 2MASS and WISE and derive the  extinction values for approximately 17 million stars. Together with the parallaxes from Gaia DR2, we then present a 3D extinction map of the southern sky for the first time.

Based on the yielded 3D extinction map, we are able to further explore the structure of the interstellar dust of the Milky
Way. Determining the structural parameters, such as the scale lengths and scale heights of the different components of the Galactic disk is one of the fundamental tasks of the Galactic studies \citep[e.g.,][]{Bland2016, Chen2017MNRAS}. Hitherto, in spite of huge datasets from various surveys,  the structural parameters of the major components of the Galactic disk, including that of the dust grains, have not been well constrained yet. Some efforts have already been made to model the distribution of the interstellar dust grains in the Milky Way. With the far-IR and near-IR COBE/DIRBE data, \citet{Drimmel2001} modeled the Galactic dust distribution with an exponential disk and obtain a scale length of 2.26\,kpc and a scale height of 134.4\,pc. With a similar dataset, \citet{Misiriotis2006} derived values of 5\,kpc and 100\,pc respectively for the scale length and height of the dusty disk. \citet{Jones2011} studied the 3D dust distribution of the Milky Way from the spectroscopic data of SDSS and obtained a scale-height of 119\,pc for the Solar neighbourhood. Most recently, \citet{Li2018} presented a model of Galactic dust distribution that has a scale-length of 3,192\,pc and a scale-height of 103\,pc, based on the LAMOST spectroscopic data. All the above analyses assume that dust grains in the Milky Way are distributed in a single disk. However, based on the 3D extinction maps deduced from the LAMOST spectroscopic surveys \citep{Yuan2015} and from multi-band photometric data \citep{Green2018, Chen2019}, Yuan et al. (in preparation) and Dobashi et al. (in preparation) independently find that the Galactic dust distribution is better modeled by  two disk components, a ``thin" component of scale height of about 70\,pc and another ``thick" component of scale height of about 200\,pc.

In the current work, we combine our newly derived 3D extinction map of the southern sky with those from the literature, and produce a 3D extinction map covering the whole sky. Based on this, we then determine the structural parameters of the Milky Way dust distribution.

The outline of this article is as follows. In Section 2, we introduce the data and method to build the 3D extinction map of the southern sky, and present the results in Section 3. The resulting dust distribution is modeled in Section 4, while our conclusions are summarized in Section 5.

\section{data and method}
\subsection{Data set}

To obtain accurate extinction values for the individual stars, we combined the SMSS DR1 optical photometry with those of Gaia DR2 in the optical, and 2MASS and WISE in the IR. The SMSS DR1 dataset contains six-band ($uvgriz$) photometry complete to roughly 18\,mag in all bands and covers an area of 17,200\,deg$^2$ of the southern sky. Sources in the SMSS DR1 are respectively cross-matched with the Gaia DR2, 2MASS and WISE catalogs with a matching radius of 1.0 arcsec, using the X-Match Service provided by CDS, Strasbourg\footnote{http://cdsxmatch.u-strasbg.fr/xmatch}.
The fraction of objects with multiple matches is less than 0.01\,\%. Stars that are detected in SkyMapper $g$, $r$ and $i$ bands, Gaia $G_{\rm{BP}}$ and $G_{\rm{RP}}$ bands, 2MASS $J$, $H$ and $K_{\rm{S}}$ bands and WISE $W$1 band and have photometric errors less than 0.08\,mag in all those nine bands are selected as our sample stars. The cuts lead to a total of about 28 million stars in our catalog.

\subsection{Extinction determination for the individual stars}

To map the 3D distribution of dust extinction in the
southern sky, we first calculate the extinction values of
all the cataloged stars. We adopt a SED fitting algorithm similar to that used by   \citet{Berry2012} and \citet{Chen2014} and apply it to multi-band photometric data of the individual stars in our catalog. In doing so, we have implicitly neglected the effects of the stellar metallicity and surface gravity and assumed that the optical and near-IR colors of the stars are determined only by their effective temperatures and values of extinction. The color of a star for any two bands, $\lambda1$ and $\lambda2$, is universal function of intrinsic $(g-i)_0$ only except for the reddening effects. Thus it can be calculated as,

\begin{equation}
c_{\rm{sim}} = c_{\rm{sl}}[(g-i)_{0}] + (C_{\lambda 1}-C_{\lambda 2}) A_{\rm{r}},
\end{equation}
where $c_{\rm{sim}}$ and $c_{\rm{sl}}$ are respectively the simulated color and the intrinsic color predicted by the reference SED library for a star of a intrinsic color $(g-i)_{\rm{0}}$, and $C_{\lambda}$ = $R_{\lambda}$/$R_{r}$ the ratio of reddening coefficients of bands $\lambda$ and $r$. We describe our reference SED library in Appendix A. For the extinction law, we adopt the extinction coefficients given by \citet{Huang2019} and \citet{Chen2019}, namely $R_{\lambda}$ = 3.407, 2.685, 2.03, 0.82, 0.52, 0.35, 0.21, 3.24 and 1.91 for $g$, $r$, $i$, $J$, $H$, $K_{\rm{S}}$, $W1$, $G_{\rm{BP}}$ and $G_{\rm{RP}}$ bands, respectively. A $\chi^2$ goodness of fit procedure is applied to each star in our catalog to obtain the best-fit values of intrinsic color $(g-i)_{0}$ and extinction $A_{r}$ for this star. We define,
\begin{equation}
\chi^{2} = \frac{1}{5} \sum_{i=1}^{7}(\frac{c^{i}_{\rm{obs}} -c^{i}_{\rm{sim}}}{\sigma_{i}})^2,
\end{equation}
where $c^{i}_{\rm{obs}}$ and $c^{i}_{\rm{sim}}$ are respectively the $i$th observed and estimated colors, and $\sigma_{i}$ the combined uncertainties of the photometric and fitting errors. In this work, we adopt seven combinations of colors, $g - r$, $r - i$, $i - J$, $J - H$, $H - K_{\rm{S}}$, $K_{\rm{S}} - W1$ and $G_{\rm{BP}} - G_{\rm{RP}}$.

\subsection{Construction of 3D Extinction map}

Stellar distances are needed to trace the extinction
in 3D. In the current work, we have adopted the distance values of the individual stars from \citet{Bailer2018}, who provided distances of about one billions stars from the Gaia DR2 parallaxes. Based on the extinction and distance estimates of the individual stars, we then mapped the distribution of dust extinction in different directions of the sky at different distances. We first group the stars into small spatial pixels (sightlines) and using the Healpix package \citep{Gorski2005}, we adopt a variable pixel size (angular resolution) for different latitudes and a minimum of 30 stars per pixel. In this work, we adopt a angular resolution of 6.9\,arcmin for Galactic latitudes $-$10\degr $< b <$ 10\degr, 13.7\,arcmin for $-$10\degr $\leqslant b \leqslant$ $-$30\degr\ and 27.5\,arcmin for $b$ $<$ $-$30\degr.

We then fit the extinction as a function of distance for each pixel with a piecewise linear function \citep{Green2015, Chen2017b}, given by,
 
\begin{equation}
A_{r}(d) = \sum_{0}^{d}(\Delta{A_r}^{i}),
\end{equation}
where $\Delta{A_r}^{i}$ is the extinction contributed by dust grains in distance bin of index $i$. The distance is binned by step $\Delta$\ $d$ = 0.2\,kpc. An MCMC analysis \citep{Metropolis1953, Hastings1970} is performed to find the best set of $\Delta{A_r}^{i}$ that maximize the likelihood defined as,

\begin{equation}
\mathcal{L} = \prod_{n=1}^{N} \frac{1}{\sqrt{2\pi}\sigma_{n}} {\rm{exp}}(\frac{-({A_{\rm{r}}}^n-{A_{\rm{r}}}^n(d))^2}{2{\sigma_n}^2}),
\end{equation}
where ${A_r}^{n}$ and ${A_r}^{n}$($d$) are respectively the $r$-band  extinction values derived from the SED fitting in the current work and that given by Eq.\,(3) for the star of index $n$ in the pixel, $d$ is the distance to the star, and $N$ the total number of stars in the pixel. The error $\sigma_{n}$ is computed by $\sigma$ = $\sqrt{{\sigma_{A_r}}^2+ ({\frac{\sigma_{d}}{d}})^2}$, where $\sigma_{A_r}$ and $\sigma_{d}$ are respectively the uncertainties of the derived extinction and distance. The uncertainty $\Delta{A_r}^{i}$ is computed from 68$\%$ probability intervals of the marginalized probability distribution functions (PDFs) of each parameter, for the accepted value after the post-burn period in the MCMC chain.

\section{3D extinction map of the southern sky}

In total, 28,173,745 stars were used for the SED fitting. Only SED fits with $\chi^2$ $<$ 2 are accepted \citep{Berry2012, Chen2014}. This yields 19,754,934 valid stars. We further exclude stars with Gaia parallax uncertainties larger than 20$\%$. This leads to a final sample of 17,375,796 stars that is used to construct the 3D extinction map of the southern sky.  

\begin{figure*}[!htbp]
\begin{minipage}[t]{1.0\linewidth}
\centering
\includegraphics[width=2.3in]{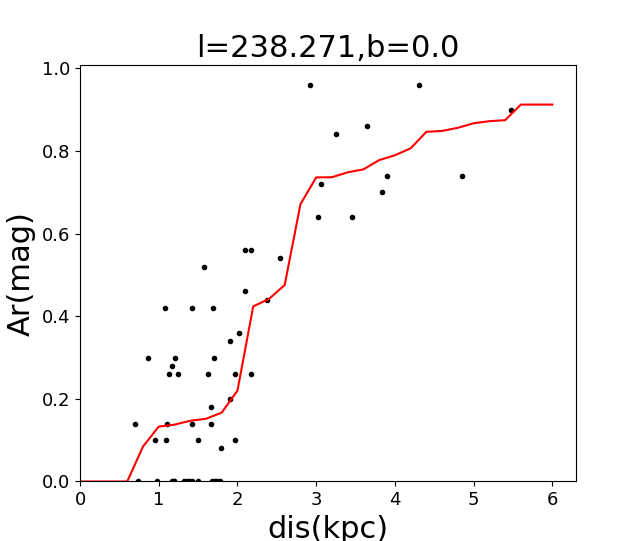}
\includegraphics[width=2.3in]{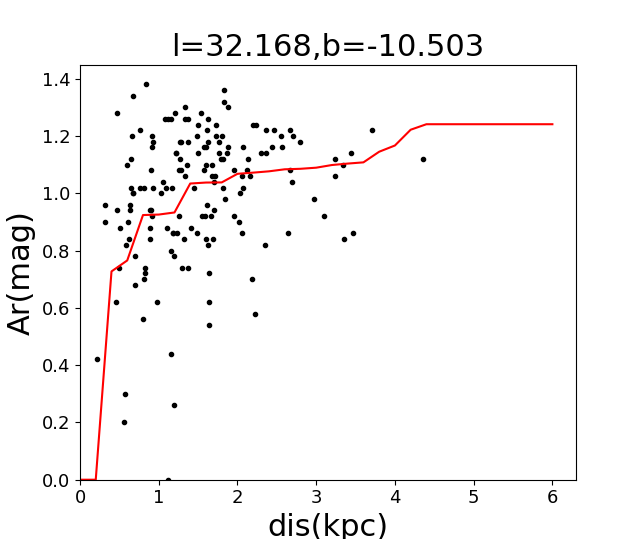}
\includegraphics[width=2.3in]{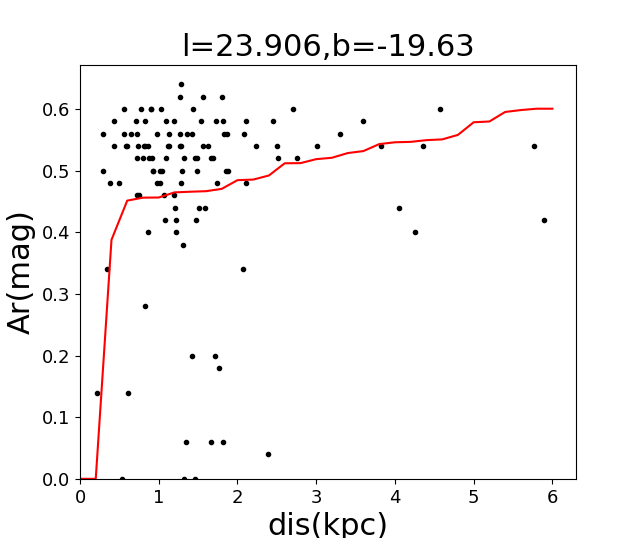}

\caption{Extinction values plotted against distances for the individual stars in three example fields. Galactic coordinates of the three pixels are respectively marked at top of the individual panels. Black dots represent $r$-band extinction and distance values of all stars and the red lines are the best-fit extinction profiles of the data.}
\end{minipage}
\label{dis_Ar}
\end{figure*}

Examples of fitting the extinction as a function of  distance for selected sight lines are shown in Fig.\,\ref{dis_Ar}. Overall, the extinction profile fits agree well with the extinction values as a function of  distance for the individual stars. Fields at low Galactic latitudes (e.g., the left panel of Fig.\,\ref{dis_Ar}) suffer from high dust extinction and the extinction keeps increasing at all distances. On the other hand, the extinction at high latitudes (e.g., the right panel of Fig.\,\ref{dis_Ar}) is much smaller and is contributed by dust grains at local distances. From the extinction profiles, one can identify the locations of molecular clouds along the sightlines. For example, we find two clouds, located  respectively at distances 0.8 and 2.2\,kpc in field ($l$, $b$) = (238.271\degr, 0.0\degr ), one cloud at distance  0.4\,kpc in filed ($l$, $b$) = (32.168\degr, $-$10.503\degr), and one cloud at distance 0.3\,kpc in field ($l$, $b$) = (23.906\degr,  $-$19.63\degr).

\begin{figure*}[!htbp]
\begin{minipage}[t]{0.5\linewidth}
\centering
\includegraphics[width=3.5in]{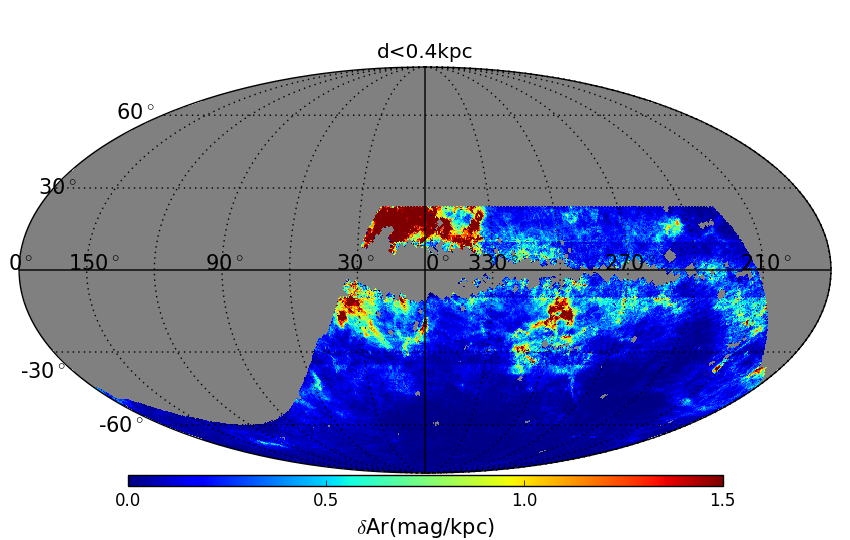}
\includegraphics[width=3.5in]{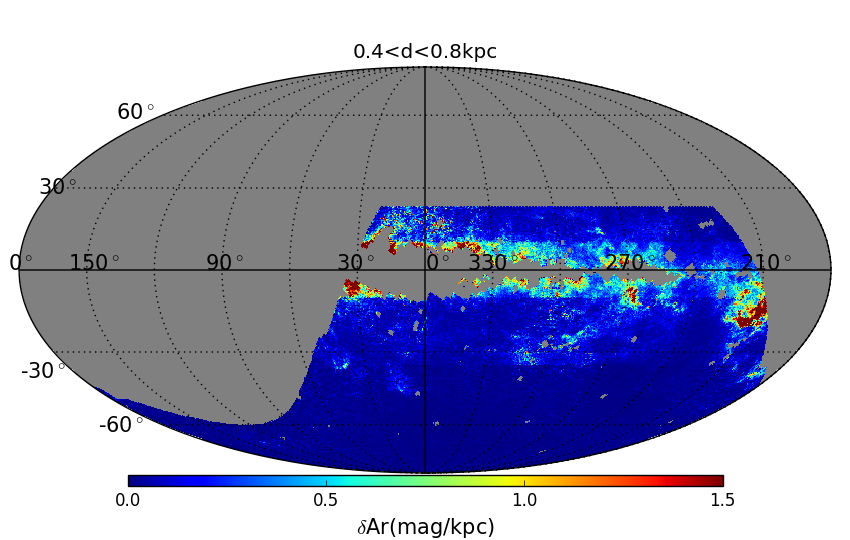}
\end{minipage}
\begin{minipage}[t]{0.5\linewidth}
\centering
\includegraphics[width=3.5in]{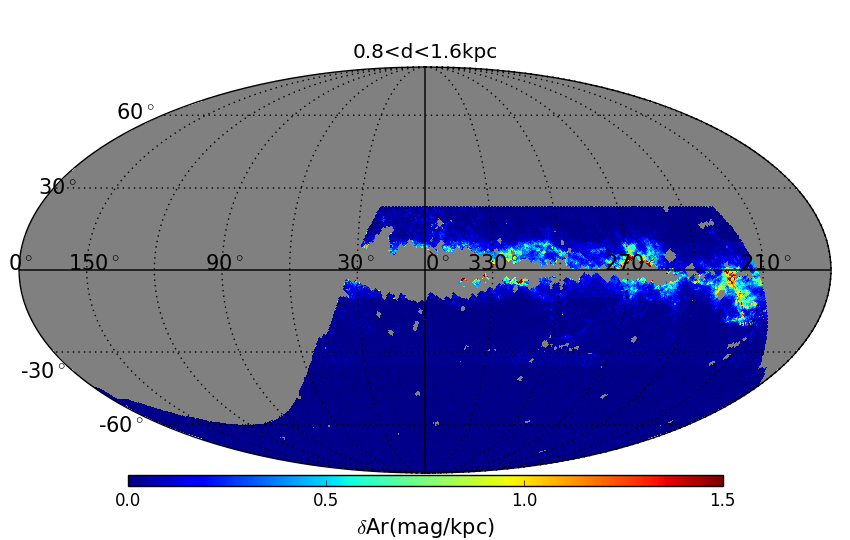}
\includegraphics[width=3.5in]{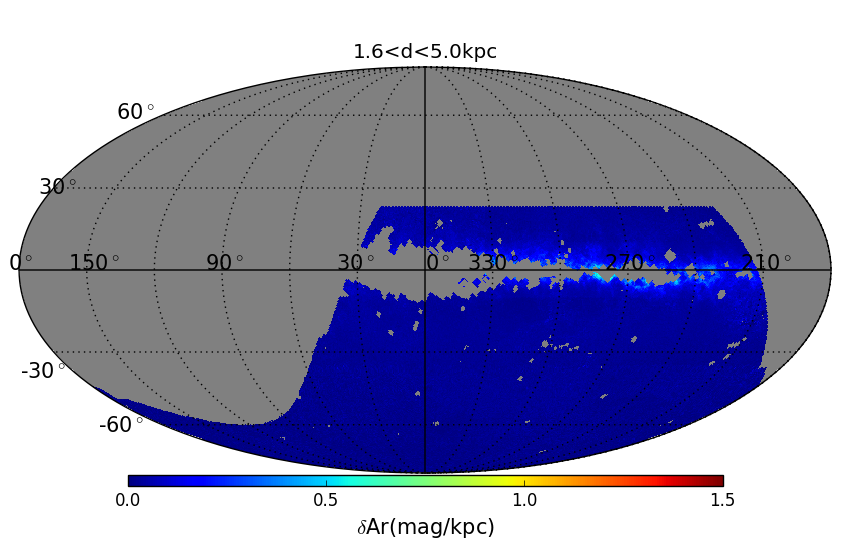}
\end{minipage}
\caption{Distributions of $r$-band extinction in the individual distance bins. The distance ranges are marked on top of the individual panels. The maps use a Galactic Mollweide projection, with $l$ = 0\degr\ at middle. Regions shaded grey are areas not covered by the current maps.}
\label{delta3dmap}
\end{figure*}
\begin{figure*}
\begin{center}
\includegraphics[width=15cm]{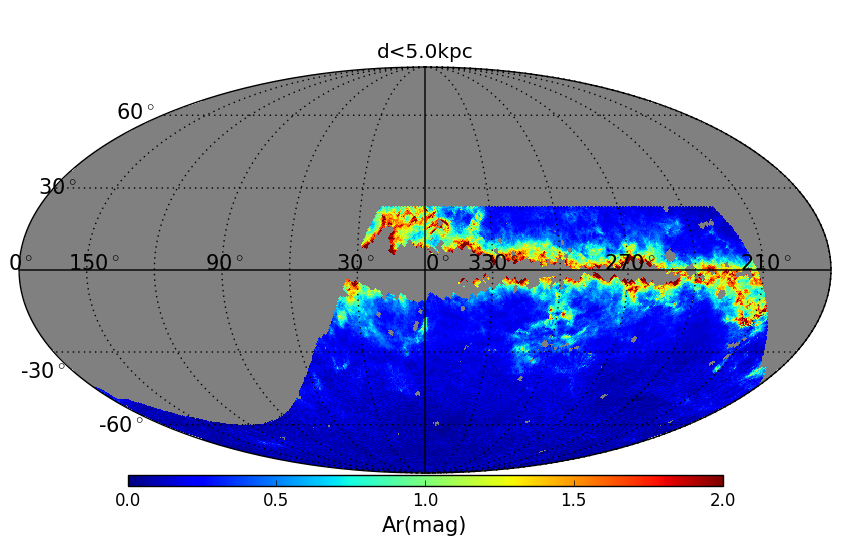}
\caption{Distribution of cumulative $r$-band extinction out to distance of 5\,kpc. The map uses a Galactic Mollweide projection, with $l$ = 0\degr\ at middle. Regions shaded grey are areas not covered by the current map.}
\end{center}
\label{3dmap}
\end{figure*}

Fig.\,\ref{delta3dmap} shows the very first set of 3D extinction maps of the southern sky that cover about 14,000 $\rm{deg}^2$ sky area at distances ranging from 0 to $\sim$ 5\,kpc.
Fig.\,\ref{delta3dmap} shows many interesting dust features. In particular, the map for the nearest distance bin ($d<$ 0.4\,kpc) shows some large-scale dust features associated with the local massive giant molecular clouds which are clearly visible in our maps, including parts of the Orion  ($l$, $b$) $\sim$ (225\degr, $-$10\degr), the Chamaeleon ($l$, $b$) $\sim$ (300\degr, $-$16\degr), the Ophiuchus ($l$, $b$) $\sim$ (355\degr, 15\degr) and the Scorpius-Centaurus-Lupus ($l$, $b$) $\sim$ (340\degr, 15\degr). Parts of Orion which extend at ($l$, $b$) $\sim$ (213\degr, $-$20\degr) and Vela ($l$, $b$) $\sim$ (270\degr, $-$5\degr) are clearly observed in our maps in the distance bin between 0.4 and 0.8\,kpc. The Monoceros at ($l$, $b$) $\sim$ (220\degr, $-$5\degr) is visible in our maps in the distance bin of 0.8 - 1.6\,kpc. These features are consistent with the distances of those clouds, $d \approx$  400, 183, 125, 140, 410 and 830\,pc for Orion, Chamaeleon, Ophiuchus, Scorpius-Centaurus-Lupus, Vela and Monoceros, respectively \citep{Zeeuw1999, Schlafly2014, Zari2018, Zucker2019, Chen2020}. The empty patches close the plane of $b$ $\sim$ 0\degr\ are due to the lack of data in the SMSS DR1. Fig.\,\ref{3dmap} shows the integrated 2D $r$-band extinction map across the footprint of the SMSS DR1. The main features agree well with the  previously published 2D extinction maps, including those from SFD and \citet{Planck2014}.

\subsection{Accessing the data}

The resultant 3D extinction maps can be accessed at {\url{https://nadc.china-vo.org/article/20200722160959?id=101032}}. In the website, we also provide a simple {\sc python} procedure, which returns reddening values for a given 3D position ($l, b$ and $d$); a simple example of how to use the procedure is given. The catalog containing the best-fit values of $A_{r}$ from the SED fitting and the distances from \citet{Bailer2018} for over 17 million stars, is also publicly available at the website.

\subsection{Comparison of extinction values of individual stars in other maps}

\begin{figure*}[!htbp]
\begin{minipage}[t]{0.5\linewidth}
\centering
\includegraphics[width=3.5in]{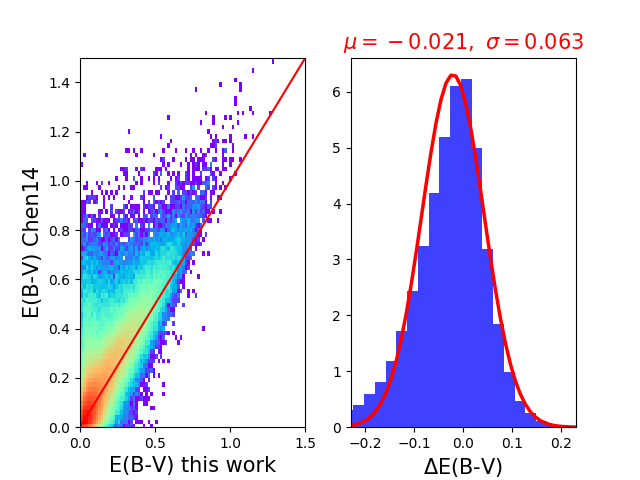}
\includegraphics[width=3.5in]{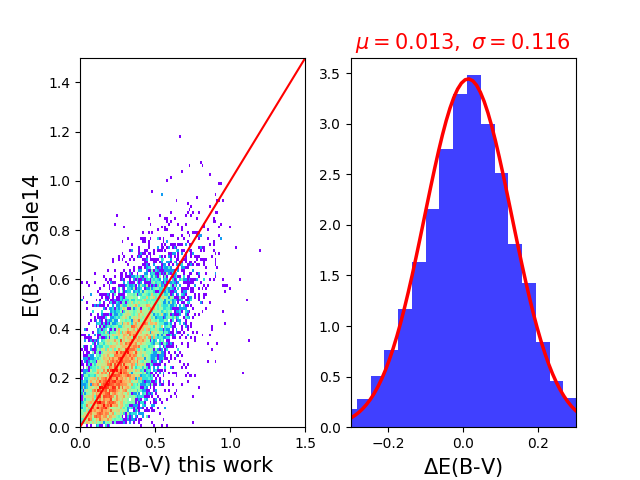}
\end{minipage}
\begin{minipage}[t]{0.5\linewidth}
\centering
\includegraphics[width=3.5in]{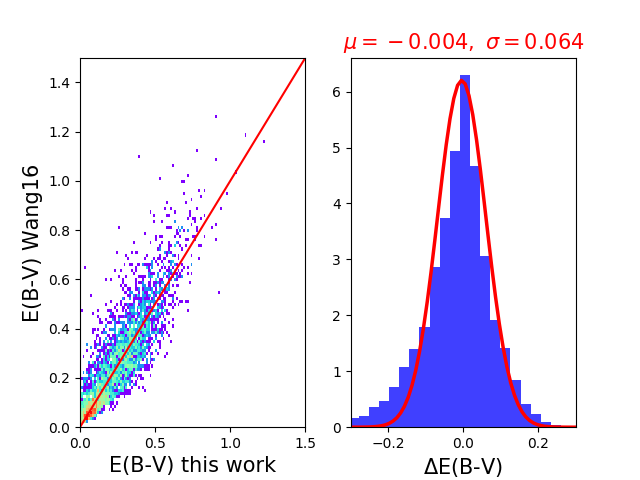}
\includegraphics[width=3.5in]{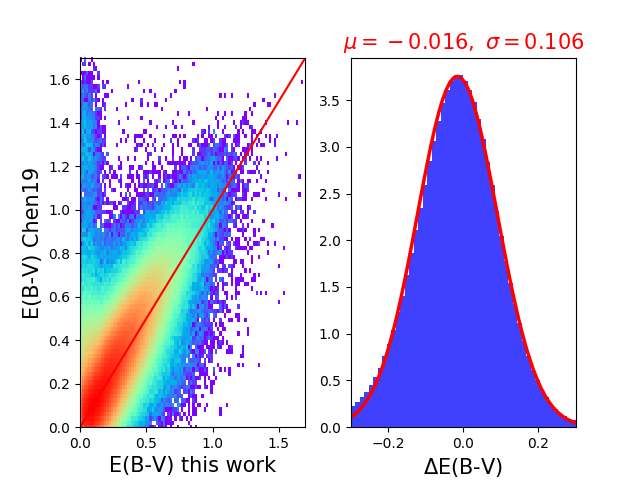}
\end{minipage}
\caption{Comparison of $E(B - V)$ values derived in the current work and those from \citet{Chen2014} (upper left), \citet{Sale2014} (bottom left), \citet{Wang2016} (upper right) and \citet{Chen2019} (bottom right). Red straight lines denoting complete equality have been plotted as reference while red curves are Gaussian fits of the distribution of differences between the maps. The corresponding means and standard deviations are on top of each histogram.}
\label{compare}
\end{figure*}

The extinction values of the individual stars presented here are compared with those obtained in the previous studies in Fig.\,\ref{compare}. Overall, our results are in good agreements with those from the literature.

\citet{Chen2014} obtained values of $r$-band extinction for over 13 million stars in the Galactic anticentre by combining the photometric measurements of XSTPS-GAC, 2MASS and WISE using a similar SED fitting algorithm. 575,651 stars analyzed by them are in common with ours. Values of $A_r$ from both works are converted to those of $E(B - V)$ using the extinction law of \citet{Yuan2013} that gives $E(B - V)$ = 0.43 $A_r$. The resulted differences have a small rms scatter of 0.063\,mag and a negligible offset of $-$0.021\,mag.

\citet{Sale2014} derived values of monochromatic extinction at 5495\,$\rm{\AA}$ of over 38 million stars in the Northern Galactic plane based on the IPHAS photometry with a hierarchical Bayesian method. The monochromatic extinction values of \citet{Sale2014} are converted to $E(B - V)$ by the relation $E(B - V)$ = 0.32 $A_0$ \citep{Chen2019}. For the 13,697 stars in common with ours, the differences between their and our results have a mean value of 0.013\,mag and an rms scatter of 0.116\,mag.

\citet{Wang2016} calculated values of $K_{\rm{S}}$-band extinction of over 0.1 million stars based on the LAMOST spectroscopic data and 2MASS photometric data using a Bayesian approach; their $A_{K_{\rm{S}}}$ values have been converted to $E(B - V)$ using the extinction law of  \citet{Cardelli1989} of $R_{V} = 3.1$ that yields $E(B - V)$ = 2.77 $A_{K_{\rm{S}}}$. With 4,098 stars in common with our catalogue, the differences between our results and those of \citet{Wang2016} have produced an rms scatter of 0.064\,mag and an offset of  $-$0.004\,mag.

\citet{Chen2019} obtained values of color excesses $E(G - K_{\rm{S}})$, $E(G_{\rm{BP}} - G_{\rm{RP}})$ and $E(H - K_{\rm{S}})$ for over 56 million stars in the Galactic plane based on the multiband photometry of Gaia DR2, 2MASS and WISE with a machine-learning algorithm. We converted their $E(G_{\rm{BP}} - G_{\rm{RP}})$ values using the relation $E(B - V)$ = 0.75 $E(G_{\rm{BP}} - G_{\rm{RP}})$. There are 3,048,284 common stars between their and our samples. Our results, compared to theirs, have an average difference of $-$0.016\,mag, along with an rms scatter of 0.106\,mag. There are a small group of stars (2,317) that have much smaller extinction values in our work than in \citet{Chen2019}. They are mostly stars of very low temperatures (spectral types later than M3) based on our analysis. The values derived by \citet{Chen2019} may be problematic due to the lack of very cool stars in their training sample.

\subsection{Comparison of 3D extinction maps with our results}

Here we compare our 3D extinction maps with the most recent ones of \citet{Green2018} and \citet{Chen2019}. 

\begin{figure*}
\begin{center}
\includegraphics[width=15cm]{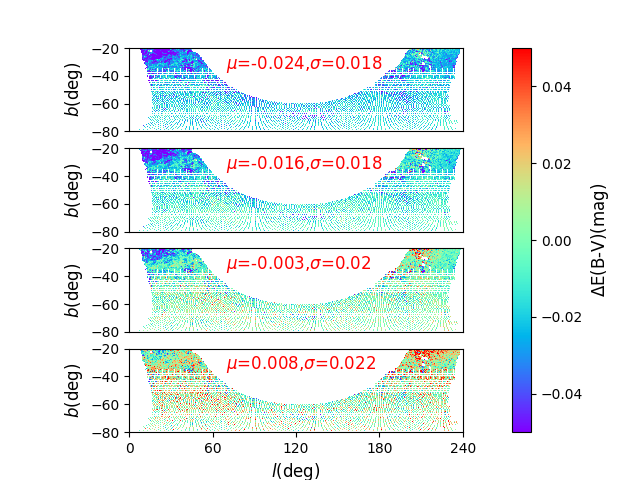}
\caption{Comparison of our resultant maps with those of \citet{Green2018}. The panels show the residuals of cumulative values of $E(B - V)$, integrated out to, from top to bottom, 1.0, 2.0, 3.0, and 5.0\,kpc. Marked in the individual panels are the means and standard deviations of the differences.}
\end{center}
\label{Skygreen}
\end{figure*}

The 3D extinction maps of \citet{Green2018} cover three quarters of the sky ($\delta$ $>$ $-$30\degr) and have angular resolutions ranging from $3.4^{\prime}$ to $13.7^{\prime}$. We converted the maps of \citet{Green2018} to match our distance resolution ($\Delta d$ = 0.2\,kpc) by linear interpolation for each sightline. In Fig.\,\ref{Skygreen} we show the comparison of the cumulative values of $E(B - V)$ yielded by our maps with those of Green et al., integrated out to distances of 1.0, 2.0, 3.0, and 5.0\,kpc, respectively. Overall, the agreement is very good, with average differences of only $\sim$ $-$0.012\,mag, and dispersions of only $\sim$ 0.02\,mag.

\begin{figure*}
\begin{center}
\includegraphics[width=15cm]{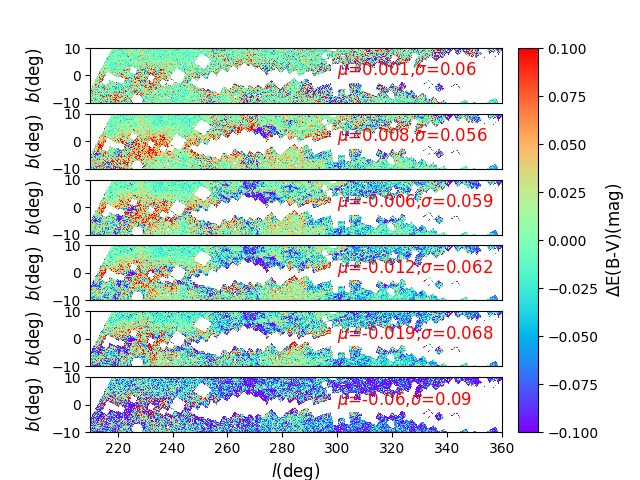}
\caption{Comparison of our resultant maps with those of \citet{Chen2019}. The panels show the residuals of cumulative values of $E(B - V)$, integrated out to, from top to bottom, 0.4, 1.0, 1.6, 2.2, 3.0 and 5.0\,kpc. Marked in the individual panels are the means and standard deviations of the differences.}
\end{center}
\label{Skychen}
\end{figure*}

The 3D extinction maps of \citet{Chen2019} cover the entire Galactic plane (0\degr $<l<$ 360\degr, $-$10\degr $<b<$ 10\degr) in three colors, $E(G - K_{\rm{S}})$, $E(G_{\rm{BP}} - G_{\rm{RP}})$ and $E(H - K_{\rm{S}})$ with a spatial angular resolution of $6^{\prime}$. We converted the maps of Chen et al. to $E(B - V)$ using the relation  $E(B - V)$ = 0.75 $E(G_{\rm{BP}} - G_{\rm{RP}})$ \citep{Chen2019}. In Fig.\,\ref{Skychen} we show the comparison of the cumulative values of $E(B - V)$ yielded by our results with those of Chen et al., integrated up to distances of 0.4, 1.0, 1.6, 2.2, 3.0 and 5.0\,kpc, respectively, for the common areas. Overall, the two sets of results agree well, with average differences of $\sim$ $-$0.01\,mag, and dispersion of $\sim$ 0.06\,mag.

\subsection{Extinction of open clusters}

We have tested our results using two well studied open clusters, NGC 2516 at $l=$ 274\degr\ and $b=-$16\degr\ with $A_r$ = 0.32\,mag (\citealt{Terndrup2002}, assuming $R_{r}=$ 2.685) and distance $d$ = 409\,pc \citep{Cantat2019} and IC 2602 at $l=$ 289.4\degr\ and $b=-$5.0\degr\ with $A_r$ = 0.183\,mag \citep{Bravi2018} and distance $d$ = 151.8\,pc \citep{Cantat2019}.

Members of NGC 2516 are selected by matching our resultant sample with the catalog of \citet{Jeffries2001}. We select sources with the flag ``F4" $=$ 1 and obtain 361 stars. The sources of IC 2602 are selected by the cross-match between our sample with \citet{Cantat2019}. Stars with membership probability of 100\% are selected. It yields 57 sources.

\begin{figure}
\centering
\includegraphics[width=0.48\textwidth]{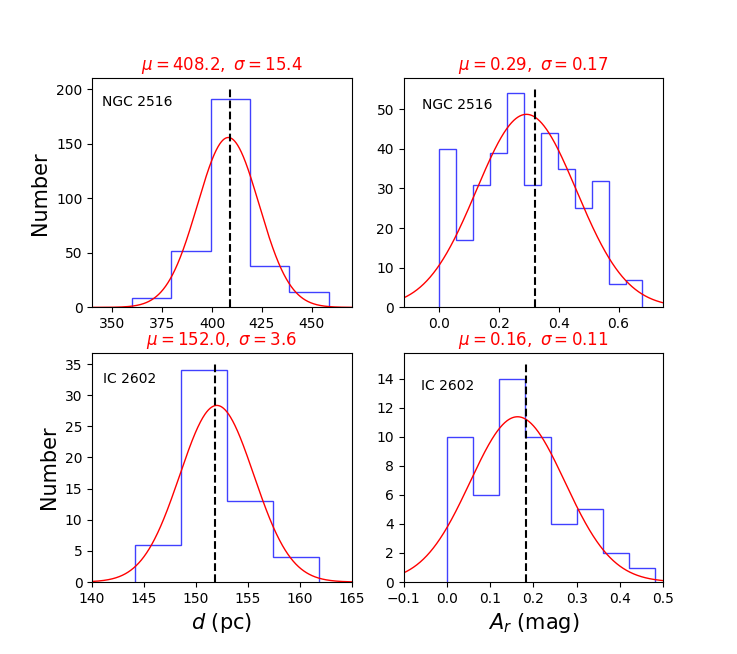}
\caption{The distribution of distance $d$ (left panels) and extinction values $A_r$ (right panels) of the member stars of NGC 2516 (upper panels) and IC 2602 (bottom panels) in our resultant sample. Red curves are Gaussian fits of the distribution. The means and dispersions are labelled on the top of each panel. The vertical dashed lines mark the reference distance/extinction values from the literature.}
\label{cluster}
\end{figure}

In Fig.\,\ref{cluster}, we show the distributions of distance and extinction estimates of the individual member stars in our sample for the two clusters. The distribution are well fitted by the Gaussian fits. We obtain $d=$ 408$\pm$15\,pc and $A_r$ = 0.29$\pm$0.17\,mag for NGC 2516 and $d=$ 152$\pm$4\,pc and $A_r$ = 0.16$\pm$0.11\,mag for IC 2602, which are in good agreement with the literature values.

\subsection{Application of the 3D extinction maps}

\begin{figure*}[!htbp]
\centering
\includegraphics[width=2.5in]{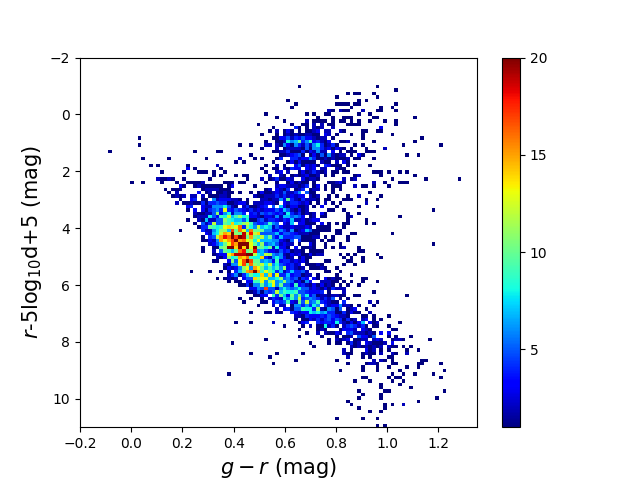}
\includegraphics[width=2.5in]{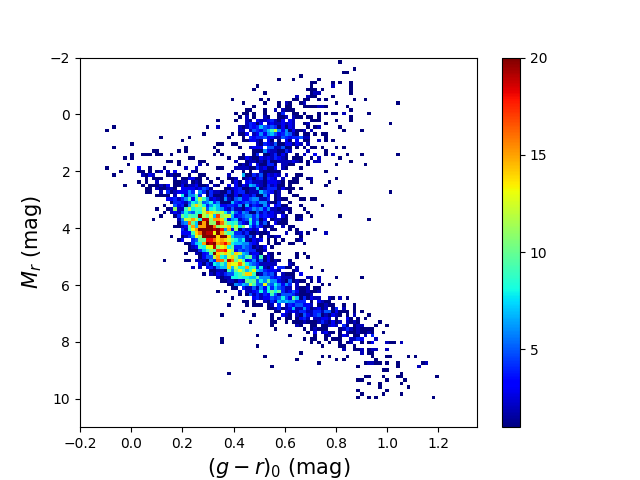}
\includegraphics[width=2.5in]{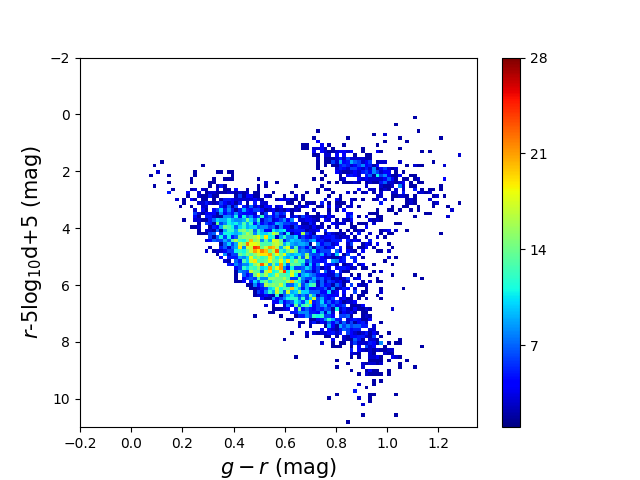}
\includegraphics[width=2.5in]{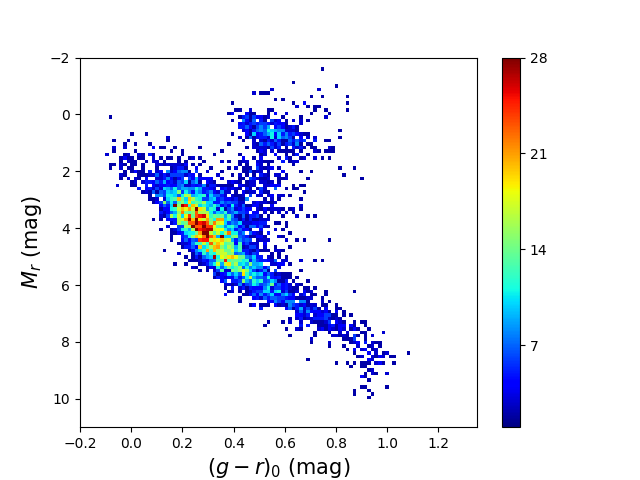}
\caption{Dust-uncorrected (left panels) and dust-corrected (right panels) color and absolute magnitude diagrams of all stars
in two example fields, one at $300\degr<l<302\degr$ and  $-17\degr<b<-15\degr$ (upper panels) and the other at $265\degr<l<267\degr$ and $-8\degr<l<-6\degr$ (bottom panels). The color scales represent the number of stars per color and magnitude bin.}
\label{CMD}
\end{figure*}

Finally, we show a simple application of our 3D extinction maps. As an example, we have applied our 3D extinction maps to correct the dust extinction of stars in two 2\degr$\times$2\degr\  fields, one centred at $l=$ 301\degr\ and $b= -$16\degr\ and the other at $l=$ 266\degr\ and $b= -$7\degr. We have selected $\sim$ 6,600 and 11,000 stars with Gaia parallax uncertainties smaller than 20\,\% for the two selected fields, respectively. The {\sc python} procedure mentioned in Sect.\,3.1 is adopted to derive the $r$-band extinction values $A_r$ for the individual stars. In Fig.\,\ref{CMD}, we compare the resultant dust-corrected color and absolute magnitude diagram of the two fields with those dust-uncorrected ones. The dust-corrected color and absolute magnitude diagrams of the two fields are in good agreement with each other. The main sequence, the red clump and the red giant branch are clearly visible. From the diagrams, we are able to obtain that the  intrinsic color and absolute magnitude of the red clump stars in the SkyMapper filters are respectively $(g-r)_0 \sim$ 0.55\,mag and $M_r \sim $ 0.6\,mag, which agree well with the theoretic values from the PARSEC isochrone\footnote{\url{http://stev.oapd.inaf.it/cgi-bin/cmd}} \citep{Salaris2002, Bressan2012}. These result indicate that our 3D extinction maps are reliable.

\
\section{THE GlOBAL DISTRIBUTION OF THE MILKY WAY DUST}

\begin{figure*}
\begin{center}
\includegraphics [width=17.0cm] {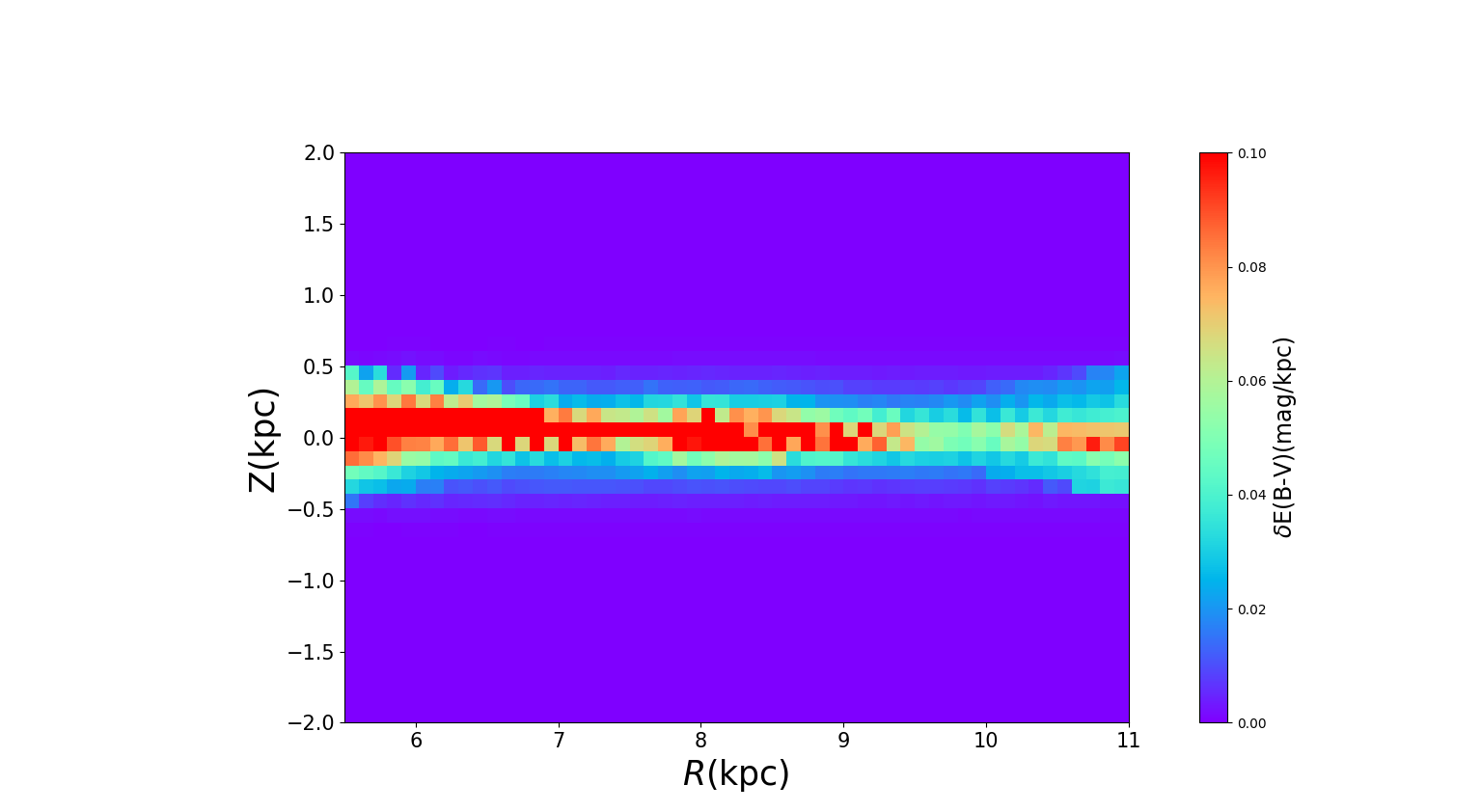}
\caption{Average $\delta E(B - V)$ (in units of mag\,kpc$^{-1}$) distribution in the $R$ and $Z$ plane. The bin size is 100\,pc $\times$ 100\,pc.}
\end{center}
\label{R_Z}
\end{figure*}

In this section, we explore the global distribution of Galactic dust grains and estimate the parameters such as the scale height. By combining the maps presented in the current work and those published by  \citet{Chen2019} and \citet{Green2018}, we obtain a 3D extinction distribution that covers the entire sky (see Appendix B). For the all-sky 3D extinction distribution, we first convert Galactic coordinates ($l$, $b$), and distance ($d$) to Cartesian coordinates $X$, $Y$ and $Z$. Here, we assum the Sun is at $R_{\odot}$ = 8.3\,kpc \citep{Reid2014}. Given the limited depth of the Gaia DR2 parallax measurements, both the 3D extinction maps of the current work and Chen et al. are only complete to $\sim$ 3\,kpc. We therefore restrict our current analysis to data points of distances between $\sim$ 6.0 and $\sim$ 11.0\,kpc, to avoid any selection bias. In Fig.\,\ref{R_Z}, we show the distribution of the average dust reddening in the $R$-$Z$ space. It shows a simple structure similar to the distribution of Galactic disk stars \citep[see Fig.\,12 of][]{Xiang2018}.

We divide the all-sky 3D extinction distribution into 9 distance bins of $R$ ranging from 6.3\,kpc to 10.8\,kpc with a bin-size of 0.5\,kpc. For each bin, the vertical distribution of the extinction is described by a second order hyperbolic function sech$^2$, as used to model the Galactic stellar disk,
\begin{equation}
\rho(Z) = \sum_{j=1}^{N}\rho_{0,j} {\rm{sech}}^2(-\frac{|Z-Z_\odot|}{H_j})+C,
\end{equation}
where $\rho_{0,j}$ and $H_j$ are dust density at the Galactic plane and scale height of the $j$th disk component. $C$ is a constant offset. In the current work, we have tried two models, one consisting of a single disk  ($N$=1) and an other consisting of two disks ($N$=2). From dust density $\rho$, the extinction is given by the standard relation,
\begin{equation}
A_\lambda = 1.086 \int_{0}^{s}\kappa_\lambda \rho ds,
\end{equation}
where $\kappa_\lambda$ is absorption coefficient. For each  bin, the average extinction as a function of $Z$ is fitted with Eq.\,(5) and Eq.\,(6) described above with an MCMC algorithm. The best-fit parameters are taken as those that give the the maximum likelihood defined as,
\begin{equation}
\mathcal{L} = \prod_{n=1}^{N} \frac{1}{\sqrt{2\pi} \sigma_{i}} {\rm{exp}}(\frac{-({E_{\rm{obs}}}^{i}-{E_{\rm{mod}}}^i)^2}{2{\sigma_{i}}^2}),
\end{equation}
where $\sigma_i$ is standard error of average extinction  $E_{\rm{obs}}^i$. The uncertainties of the resultant parameters are derived again using a Monte Carlo method. For each bin, we randomly generate 1000 samples accounting for uncertainties in their reddening estimates. The MCMC analysis is then applied to each sample and the resulted best-fit parameters are recorded. Gaussian distributions are expected for the derived parameters. We take the dispersions of the Gaussian distributions as the errors of the corresponding parameters.

\begin{figure*}[!htbp]
\begin{minipage}[t]{0.5\linewidth}
\centering
\includegraphics[width=3.5in]{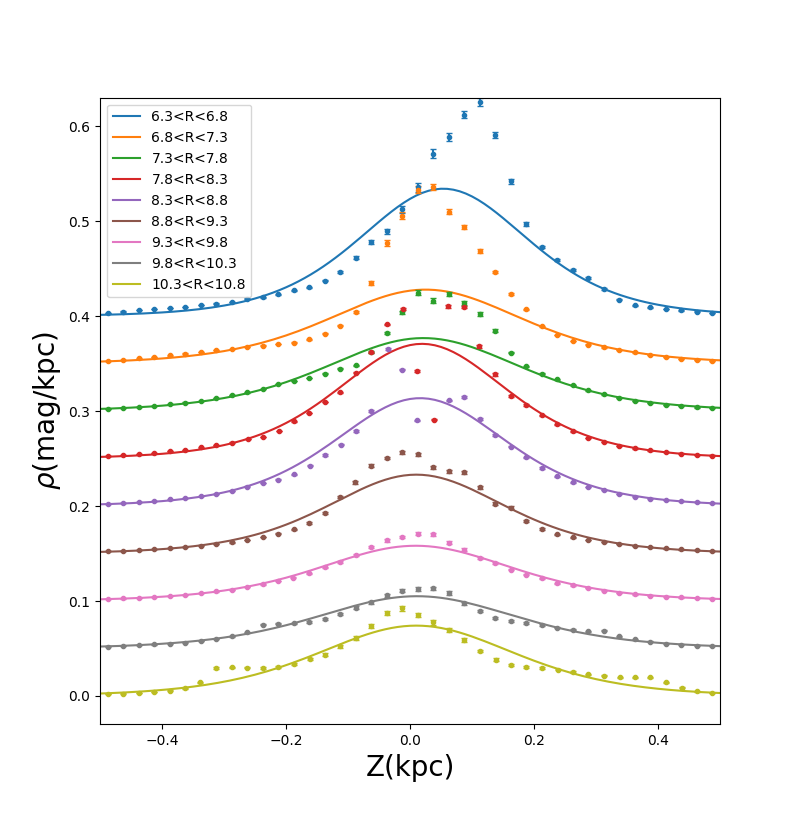}
\end{minipage}
\begin{minipage}[t]{0.5\linewidth}
\centering
\includegraphics[width=3.5in]{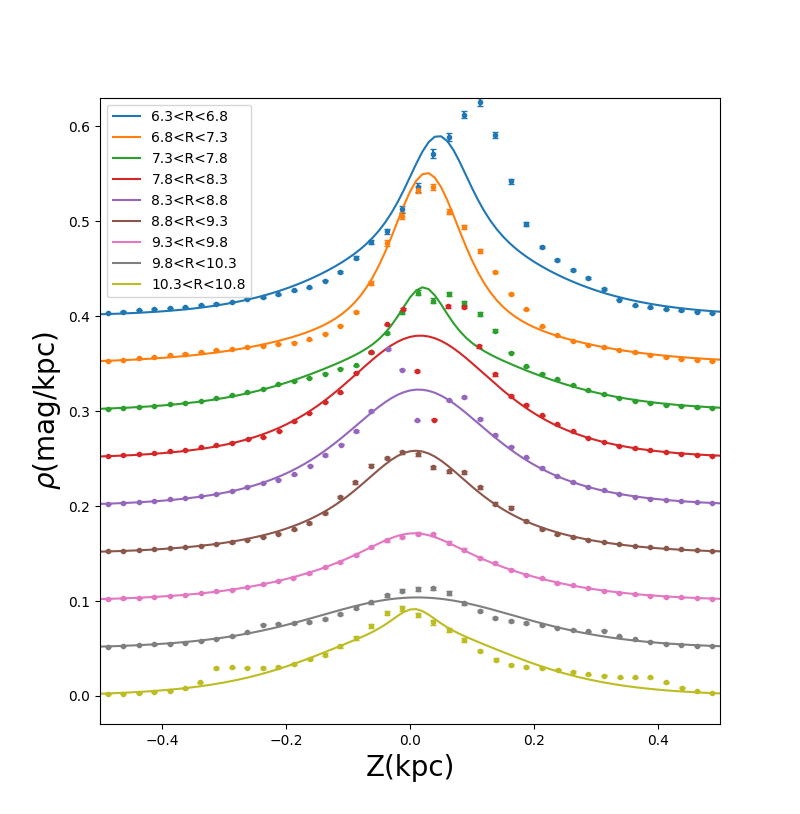}
\end{minipage}
\caption{Fit the vertical dust distributions with respectively a single-disk model (left panel) and a two-disk model (right panel) in the different $R$ slices. The black dots and error bars are median and standard errors of the mean values in bins of $Z$ of 25\,pc width. Curves of different colors are the fits.}
\label{niheh}
\end{figure*}

\begin{figure*}[!htbp]
\begin{minipage}[t]{0.5\linewidth}
\centering
\includegraphics[width=3.5in]{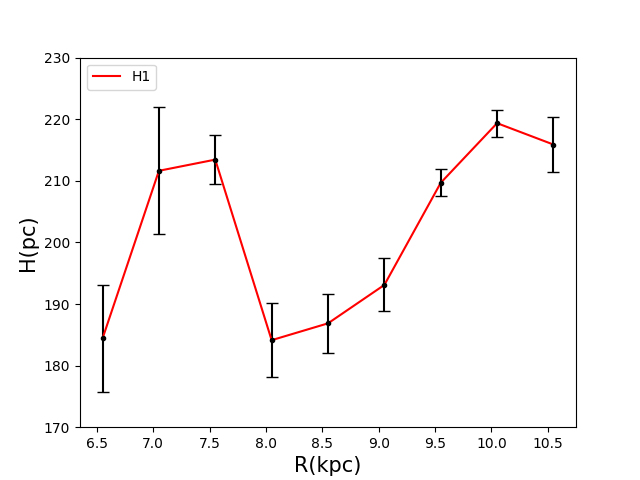}
\end{minipage}
\begin{minipage}[t]{0.5\linewidth}
\centering
\includegraphics[width=3.5in]{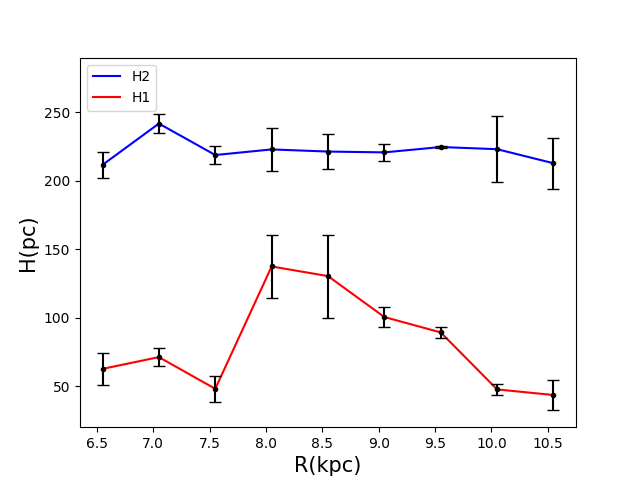}
\end{minipage}
\caption{Resultant scale-height $H_{j}$ as a function of the radial distance $R$ for the single-disk model (left panel) and the two-disk model (right panel) respectively.}
\label{R_H}
\end{figure*}

Fig.\,\ref{niheh} shows the fit results. The resulted parameters are listed in Table \ref{para}. The resulted scale-heights are plotted against  radial distance $R$ in Fig.\,\ref{R_H}. For the single-disk model, we obtain an average scale-height of 205.5$\pm$1.5\,pc. This value is slightly larger than reported in the literature \citep{Li2018, Jones2011, Drimmel2001}. This is mainly caused by the poor fitting in the current work. For the two-disk model, the resulted average scale-height for the first (``thin") disk is 72.7$\pm$2.2 pc and that for the second (``thick") disk is 224.6$\pm$0.7\,pc. Our results for the two-disk model are consistent with those of Yuan et al. (in preparation) and Dobashi et al. (in preparation). Given the relatively large uncertainties and the limited $R$ ranges, it is difficult to tell at the moment that if there are any correlation between the resulted scale heights and the radial distances.

Fig.\,\ref{niheh} clearly shows that the two-disk model fits the data much better than the single-disk model. To discriminate between these two models, we introduce the Bayesian Information Criterion \citep[BIC;][] {Schwarz1978} defined as,
\begin{equation}
{\rm{BIC}} \equiv -2{\rm{ln}}\mathcal{L}_{\rm{max}} + k{\rm{ln}}N,
\end{equation}
where $\mathcal{L}_{\rm{max}}$ is the maximum likelihood achieved by the model, $k$ the number of free parameters of the model and $N$ the number of data points used for the fit. In Table \ref{para} we list the resulted BIC values for both models for the individual bins of $R$. In all bins, the two-disk model yields significantly lower values of BIC than the one-disk model, indicating better fits.

\begin{table*}
\centering
\caption{Parameters of dust distribution.}
\tiny 
\begin{tabular}{l}
(a) Single-disk model  \\
\end{tabular}
\begin{tabular}{llllllllll}
\hline
R (kpc) & 6.3-6.8 & 6.8-7.3 & 7.3-7.8 & 7.8-8.3 & 8.3-8.8 & 8.8-9.3                                         & 9.3-9.8  & 9.8-10.3 & 10.3-10.8 \\
\hline
$\rho_{1}$   ($mag/kpc$) &  0.135$\pm$0.015 &  0.079$\pm$0.010  &  0.077$\pm$0.004 &  0.121$\pm$0.010 &  0.115$\pm$0.009  &  0.085$\pm$0.005  &  0.059$\pm$0.002 &  0.055$\pm$0.001 &  0.074$\pm$0.003  \\
$H_{1}$   ($pc$) &  184.9$\pm$8.8 &  210.4$\pm$10.3  &  213.6$\pm$3.9 &  184.2$\pm$6.0 &  186.6$\pm$4.8  &  191.3$\pm$4.3  &  208.9$\pm$2.2 &  219.4$\pm$2.2 &  216.4$\pm$4.5  \\
${Z_{\odot}}$   ($pc$) &  80.0$\pm$5.3 &  52.2$\pm$4.3  &  48.6$\pm$1.8 &  46.2$\pm$3.2 &  43.3$\pm$2.2  &  37.0$\pm$2.6  &  36.0$\pm$1.2 &  38.3$\pm$2.9 &  37.4$\pm$5.8  \\
BIC   &  5947.3 & 12513.7 &	4171.9 & 27039.8 & 7890.8 &	2111.1 &	-715.7 & -471.7 & 2219.8  \\
\hline
\end{tabular}
\begin{tabular}{l}
(b)Two-disk model  \\
\end{tabular}
\begin{tabular}{llllllllll}
\hline
R (kpc) & 6.3-6.8 & 6.8-7.3 & 7.3-7.8 & 7.8-8.3 & 8.3-8.8 & 8.8-9.3  & 9.3-9.8  & 9.8-10.3 & 10.3-10.8 \\
\hline
$\rho_{1}$   ($mag/kpc$) & 0.126$\pm$0.021 &  0.147$\pm$0.010  &  0.059$\pm$0.011 &  0.080$\pm$0.014 &  0.073$\pm$0.019  &  0.063$\pm$0.007  &  0.025$\pm$0.001 &  0.014$\pm$0.002 &  0.025$\pm$0.016  \\
$\rho_{2}$   ($mag/kpc$) &  0.082$\pm$0.011 &  0.054$\pm$0.004  &  0.072$\pm$0.005 &  0.056$\pm$0.022 &  0.055$\pm$0.018  &  0.045$\pm$0.005  &  0.047$\pm$0.001 &  0.051$\pm$0.001 &  0.068$\pm$0.015 \\
$H_{1}$   ($pc$) &  62.6$\pm$11.7 &  71.2$\pm$6.7  &  48.0$\pm$9.6 &  137.4$\pm$23.0 &  130.4$\pm$30.3  &  100.5$\pm$7.4  &  89.2$\pm$4.0 &  47.6$\pm$4.0 &  43.5$\pm$11.0  \\
$H_{2}$   ($pc$) &  211.7$\pm$9.4 &  241.8$\pm$6.9  &  218.8$\pm$6.3 &  222.9$\pm$15.6 &  221.4$\pm$12.8  &  220.8$\pm$6.3  &  224.7$\pm$0.7 &  223.1$\pm$24.1 &  212.8$\pm$18.7  \\
${Z_{\odot}}$   ($pc$) &  83.3$\pm$5.2 &  54.0$\pm$3.5  &  48.4$\pm$2.1 &  43.4$\pm$3.8 &  40.1$\pm$2.4  &  36.9$\pm$1.3  &  34.0$\pm$0.4 &  38.1$\pm$3.0 &  37.8$\pm$4.9  \\
BIC   &  2855.6 & 4995.0 & 2599.6 &	24118.3 & 5982.5 & -731.1 &	-1229.8 & -826.1 & 303.7 \\
\hline
\end{tabular}
\label{para}
\end{table*}

\section{Summary}

By combing the photometry of SMSS DR1, Gaia DR2, 2MASS and WISE, we have obtained a sample containing 28 million stars with high quality multi-band photometric measurements in the optical and near-IR. We have applied an SED fitting algorithm to the individual stars and obtained values of $r$-band extinction $A_r$ for 19 million objects. Combining with the distances estimated by \citet{Bailer2018} from the Gaia parallax measurements, we have obtained a sample of about 17 million stars with Gaia DR2 parallax uncertainties smaller than 20$\%$. Based on the sample, we have constructed a 3D extinction map for the southern sky. The map covers over 14,000 $\rm{deg}^2$ with angular resolutions between 6.9\,arcmin and 27\,arcmin up to a distance of about 5\,kpc from the Sun.

By combining our 3D extinction map with those from the literature, we have constructed an all-sky 3D extinction map and explored the vertical distribution of the interstellar dust in the Milky Way. We have adopted two different models, one consisting of a single disk and another consisting of two disks. For the single-disk model, we obtain an average scale-height of 205.5$\pm$1.5\,pc. For the two-disk model, the ``thin" disk has an  average scale-height of 72.7$\pm$2.4\,pc, while the ``thick" disk has an  average scale-height of 224.6$\pm$0.7\,pc. For all radial bins, the two-disk model fits the data significantly better than the single-disk model.

\section*{Acknowledgements}

We want to thank the anonymous referee for detailed and constructive comments that improve the manuscript significantly. This work is partially supported by National Key R\&D Program of China No.~2019YFA0405503, National Natural Science Foundation of China grants No.\,11803029, 11833006 and U1731308 and Yunnan University grant No.~C1762201\\00007. HBY is supported by NSFC grant No.~11603002 and Beijing Normal University grant No.~310232102. This research made use of the cross-match service provided by CDS, Strasbourg.

This work has made use of data products from the SkyMapper (SkyMapper Southern Sky Survey). SkyMapper is owned and operated by The Australian National University’s (ANU) Research School of Astronomy and Astrophysics (RSAA). The survey data were processed and provided by the SkyMapper Team at ANU. The SkyMapper website is {\url{http://skymapper.anu.edu.au}}

This work presents results from the European Space Agency (ESA) space mission Gaia. Gaia data are being processed by the Gaia Data Processing and Analysis Consortium (DPAC). Funding for the DPAC is provided by national institutions, in particular the institutions participating in the Gaia MultiLateral Agreement (MLA). The Gaia mission website is {\url{https://www.cosmos.esa.int/gaia}}. The Gaia archive website is {\url{https://archives.esac.esa.int/gaia}}.

This publication makes use of data products from the Two Micron All Sky Survey, which is a joint project of the University of Massachusetts and the Infrared Processing and Analysis Center/California Institute of Technology, funded by the National Aeronautics and Space Administration and the National Science Foundation.

This publication makes use of data products from the Widefield Infrared Survey Explorer, which is a joint project of the University of California, Los Angeles, and the Jet Propulsion Laboratory/California Institute of Technology, funded by the National Aeronautics and Space Administration.



\bibliographystyle{apj}
\bibliography{obsamp}



\begin{appendix}

\section{Reference stellar locus}
\begin{figure*}
\begin{center}
\includegraphics [width=18cm,,height=20cm] {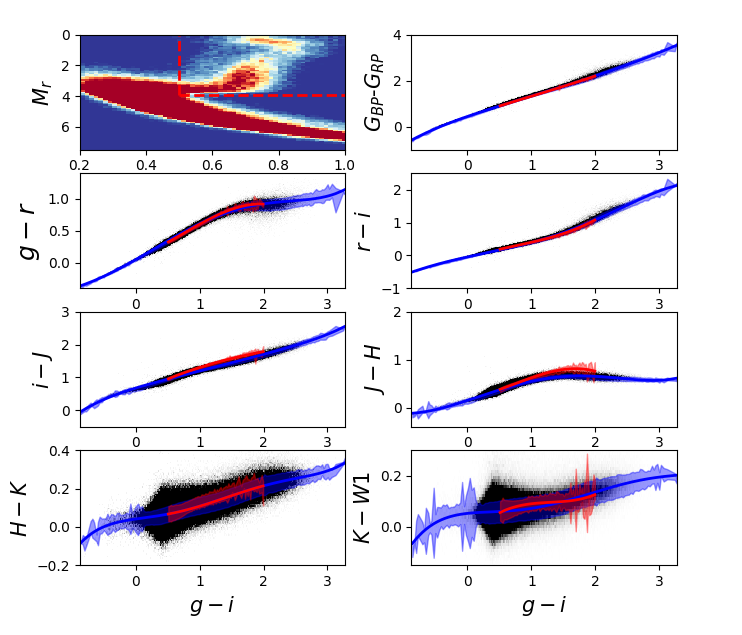}
\caption{Fits of stellar loci in the SkyMapper, Gaia, 2MASS and WISE color planes. Approximately 1 million stars  are used to obtain the fits. The fitted stellar loci are over plotted on stellar number density distributions in the individual color-color planes. Black and red filled shadows represent respectively the standard deviations of the colors for dwarfs and giants.}
\end{center}
\label{loci}
\end{figure*}

In this work, we have applied spectral energy distribution (SED) fitting to multi-band photometric data. The method  requires a library of reference stellar loci. To build the library, we first select 1,134,293 stars of $E(B-V)_{\rm{SFD}}$ $<$ 0.02\,mag and detected in all 9 bands of SkyMapper $g$, $r$ and $i$, Gaia $G_{\rm{BP}}$ and  $G_{\rm{RP}}$, 2MASS $J$, $H$ and $K_{\rm{S}}$, and WISE $W1$, with photometric errors less than 0.05 mag in all the bands. Secondly, based on the color magnitude diagram (CMD) (see the first panel of Fig.\,\ref{loci}), we divide our sample stars into two classes, a dwarf sub-sample containing 987,336 dwarfs that is used to build the stellar loci of dwarfs, and a sub-sample containing 146,957 giants that is used to build the stellar loci for giants\footnote{The two samples can be accessed from \url{https://nadc.china-vo.org/article/20200722160959?id=101032}}. After corrected for the effects of reddening, we then, for each star, create a stellar locus in seven color $g - r$, $r - i$, $i - J$, $J - H$, $H - K_{\rm{S}}$, $K_{\rm{S}} - W1$ and $G_{\rm{BP}} - G_{\rm{RP}}$. The locus is labeled by intrinsic color $(g -i)_{0}$. To fit the stellar loci, we use fifth order and third order polynomials for dwarfs and giants, respectively. The resulted stellar loci are plotted in Fig.\,\ref{loci}. The fit parameters are presented in Table\,\ref{parameter}. The adopted polynomial functions fit the data very well, with typical median residuals of $\sim$ 0.01\,mag for dwarfs and $\sim$ 0.004\,mag for giants, respectively. The resultant fits are used to generate reference stellar loci  adopted in the current work.

\section{A THREE-DIMENSIONAL EXTINCTION MAP OF THE ENTIRE SKY}

By combining the 3D extinction map presented in the current work and those published previously by \citep{Chen2019, Green2018}, we have obtained a 3D extinction map that covers the entire sky. The 3D map resented in this work covers the SMSS area of $\sim$ 14,000\,${\rm deg^{2}}$ and has spatial resolutions between 6.9 and 27\,arcmin. The map of \citet[][based on data of the Gaia DR2, 2MASS and WISE]{Chen2019} covers the area of the Galactic plane (0\degr $< l <$ 360\degr, $-$10\degr $< b <$ 10\degr), whereas that of \citet[][based on data of the Pan-STARRS and 2MASS]{Green2018} covers three quarters of the sky (Declination $\delta$ $>$ $-$30\degr). The all-sky 3D extinction map has been build using 860 million stars divided into 4.42 million sightlines.

\begin{figure*}
\begin{center}
\includegraphics[width=15cm]{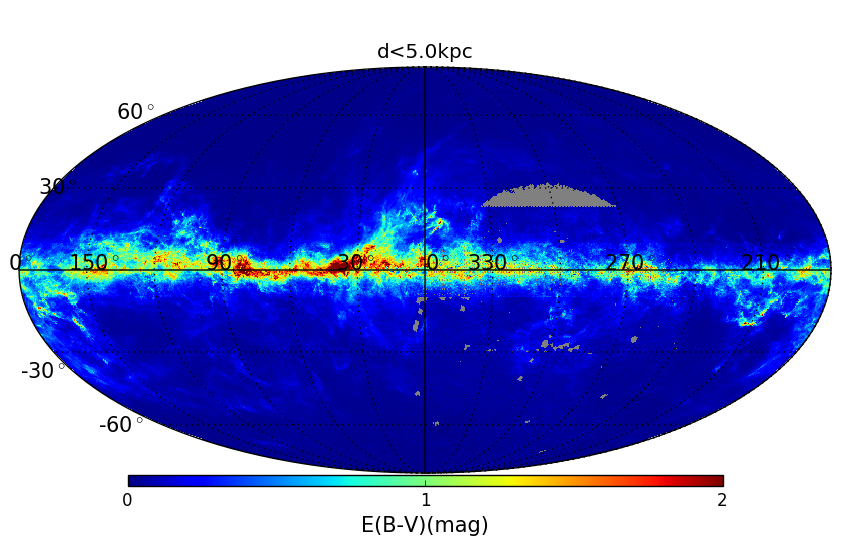}
\caption{Distribution of cumulative extinction in $E(B - V)$ integrated to distance 5\,kpc. The map uses a Galactic Mollweide projection, with $l$ = 0\degr\ at middle. Regions shaded grey are areas not covered by our map.}
\end{center}
\label{allsky5}
\end{figure*}

\begin{figure*}
\begin{center}
\includegraphics[width=15cm]{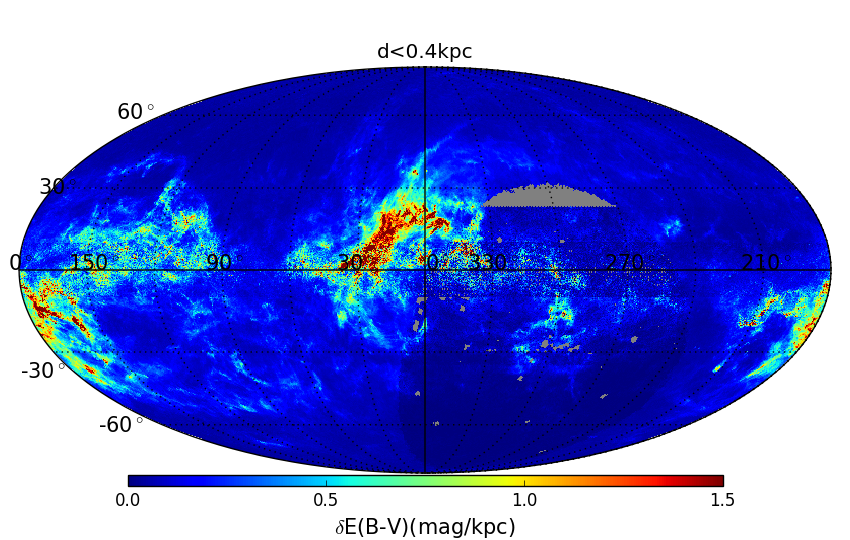}
\caption{Distribution of differential extinction in $\delta E(B - V)$ within distance $d <$ 0.4\,kpc. The map uses a Galactic Mollweide projection, with $l$ = 0\degr\ at middle. Regions shaded grey are areas not covered by our map.}
\end{center}
\label{allsky0.4}
\end{figure*}

Specifically, we follow the hierarchical HEALPix format of
\citet{Green2018}. The resulted map for the whole sky has high angular resolutions, typically ranging between $3.4^{\prime}$ and $13.7^{\prime}$. The map of \citet{Chen2019} is on a grid of HEALPix pixels with $N_{\rm{side}}$ = 512 or a spatial angular resolution of $6^{\prime}$. Our current map for the southern sky of high latitudes ($b$ $<$ $-$30\degr) is on a grid of HEALPix pixels with $N_{\rm{side}}$ = 128. For the whole sky map, we eventually employ an angular resolution of 6.9\,arcmin for Galactic latitude $|b|$ $<$ 10\degr\ ($N_{\rm{side}}$ = 512), 14\,arcmin ($N_{\rm{side}}$ = 256) for 10\degr $\leqslant |b| \leqslant$ 30\degr\ and 27\,arcmin ($N_{\rm{side}}$ = 128) for $|b|$ $>$ 30\degr. The all-sky 3D extinction map is publicly available on our website (see Sect.~3.1). The resulted all-sky map of $E(B - V)$ integrated to distance 5\,kpc is displayed in Fig.\,\ref{allsky5}, whereas extinction map of $\delta E(B - V)$ at $d$ $<$ 0.4\,kpc is displayed in Fig.\,\ref{allsky0.4}. In Fig.\,\ref{allsky5}, we recover the familiar dust structures, such as the Taurus and Perseus at $l$ $>$ 140\degr\ and below the Galactic plane; the Cepheus at $l$ $>$ 100\degr\ and above the Galactic plane; the Aquila rift at ($l$, $b$) $\sim$ (30\degr , +3\degr ); the Ophiuchus at ($l$, $b$) $\sim$ (350\degr , +15\degr ); the Scorpius-Centaurus-Lupus at 300 $<$ $l$ $<$ 350\degr, slightly above the Galactic plane; the Chamaeleon, at ($l$, $b$) $\sim$ (300\degr , $-$16\degr) and the Vela at 240\degr $<$ $l$ $<$ 270\degr.

\begin{table*}
\centering
\caption{Coefficients of polynomial fits to the stellar loci}
\begin{tabular}{ccccccccc}
\hline
\hline
Color & $A_{0}$ & $A_{1}$ & $A_{2}$ & $A_{3}$ & $A_{4}$ & $A_{5}$ & Maximum residual & Median residual \\
\hline
 Dwarfs \\
$G_{\rm{BP}}-G_{\rm{RP}}$ & $0.4485$ & $0.9593$& $-0.1549$& $0.091$& $-0.0208$& $0.002$& $0.2469$& $0.0155$\\
$g-r$ & $0.0479$ & $0.574$& $0.0608$& $-0.073$& $-0.005$& $0.0047$& $0.2623$& $0.0121$\\
$r-i$ & $-0.0477$ & $0.4257$& $-0.0632$& $0.0763$& $0.0039$& $-0.0045$& $0.2614$& $0.0124$\\
$i-J$ & $0.6773$ & $0.5962$& $-0.1541$& $0.1084$& $-0.0361$& $0.005$& $0.2198$& $0.0195$\\
$J-H$ & $0.1474$ & $0.4378$& $0.054$& $-0.1029$& $0.0095$& $0.0026$& $0.0611$& $0.0139$\\
$H-K$ & $0.041$ & $0.048$& $-0.0216$& $0.0708$& $-0.0358$& $0.0053$& $0.0394$& $0.0037$\\
$K-W1$ & $0.0528$ & $0.0273$& $-0.0533$& $0.0651$& $-0.0221$& $0.0023$& $0.0465$& $0.0026$\\
\hline
 Giants \\
$G_{\rm{BP}}-G_{\rm{RP}}$ & $0.321$ & $1.3117$& $-0.3642$ & $0.0915$ &{} &{} & $0.0605$& $0.0012$\\
$g-r$ & $0.0989$ & $0.3057$& $0.3437$& $-0.1468$ &{} &{} & $0.0612$& $0.0036$\\
$r-i$ & $-0.0935$ & $0.6806$& $-0.3327$& $0.1441$ &{} &{} & $0.0613$& $0.0035$\\
$i-J$ & $0.471$ & $1.0966$& $-0.3106$& $0.0468$ &{} &{} & $0.0916$& $0.0039$\\
$J-H$ & $0.1497$ & $0.3442$& $0.2694$& $-0.1426$ &{} &{} & $0.1678$& $0.0051$\\
$H-K$ & $0.0395$ & $-0.0063$& $0.1404$& $-0.0461$ &{} &{} & $0.0448$& $0.0038$\\
$K-W1$ & $-0.0356$ & $0.2627$& $-0.1816$& $0.0445$ &{} &{} & $0.0373$& $0.0027$\\
\hline
\end{tabular}
\label{parameter}
\end{table*}

\end{appendix}
\end{document}